\documentclass{article}
\usepackage{spconf,amsmath,graphicx,url}

\usepackage{enumitem}
\setlist{nosep, leftmargin=14pt}
\usepackage{amssymb}
\usepackage[dvipsnames]{xcolor}

\usepackage{mwe} % to get dummy images
\usepackage{color}

\usepackage[colorlinks=true,linkcolor=blue, citecolor=blue, urlcolor=blue]{hyperref} 

\title{Novel OCT mosaicking pipeline with Feature- and Pixel-based registration}
\name{Jiacheng Wang, Hao Li, Dewei Hu, Yuankai K. Tao, Ipek Oguz} %\thanks{Some author footnote.}}
\address{Vanderbilt University}
\begin{document}
%\ninept
%
\maketitle
\begin{abstract}
 High-resolution Optical Coherence Tomography (OCT) images are crucial for ophthalmology studies but are limited by their relatively narrow field of view (FoV). Image mosaicking is a technique for aligning multiple overlapping images to obtain a larger FoV. Current mosaicking pipelines often struggle with substantial noise and considerable displacement between the input sub-fields. In this paper, we propose a versatile pipeline for stitching multi-view OCT/OCTA \textit{en face} projection images. Our method combines the strengths of learning-based feature matching and robust pixel-based registration to align multiple images effectively. Furthermore, we advance the application of a trained foundational model, Segment Anything Model (SAM), to validate mosaicking results in an unsupervised manner. The efficacy of our pipeline is validated using an in-house dataset and a large public dataset, where our method shows superior performance in terms of both accuracy and computational efficiency. We also made our evaluation tool for image mosaicking and the corresponding pipeline publicly available at \url{https://github.com/MedICL-VU/OCT-mosaicking}.

\end{abstract}
\section{Introduction}
\label{sec:intro}

Image mosaicking methods detect, align and blend geometrically overlapping regions in multiple images to generate a stitching result with a wider field of view \cite{ghosh2016survey}. %These methods are widely used in various fields of computer vision and graphics, ranging from commercial usage to scientific research. 
%There is also a growing interest in developing biomedical imaging mosaicking in clinical applications \cite{bano2024image}, including X-ray image stitching for better anatomical diagnosis \cite{adwan2016new}, whole slide mosaicking (WSI) in histopathology \cite{chalfoun2017mist}, expanded endoscopy images for improved cystoscopy and endoscopic surgeries \cite{bano2020deep}, and Optical Coherence Tomography (OCT) for ophthalmology \cite{el2018spectrally}. 
There are many applications of image mosaicking in biomedical imaging \cite{bano2024image, adwan2016new,bano2020deep,el2018spectrally}.
%Each biomedical application is rooted in its own challenges, making it still an open question to study a unified solution tackling problems ranging from efficiency to accuracy.  

There are two categories of approaches for image mosaicking. Feature-based techniques \cite{adwan2016new, yang2018multi} seek feature point correspondences between images, enriched with geometric and content descriptors. 
%Then proper geometric estimation \cite{fischler1981random}, and blending technique was adopted. 
In contrast, pixel-based registration methods strive for direct pixel intensity similarity across images \cite{keszei2017survey}, employing various cost functions and iterative optimization. However, both methods have limitations, especially when applied in medical image contexts. Feature-based methods can be affected by the quality of content, such as noise levels, reflections, and motion blur \cite{bano2020deep}. On the other hand, pixel-based methods have a limited convergence range and require extensive overlap between the images. This necessity makes them less robust for images with limited overlap, as they are prone to fall into local optima \cite{bano2024image}.

\begin{figure*}[ht]
\centering
\includegraphics[width=1\linewidth]{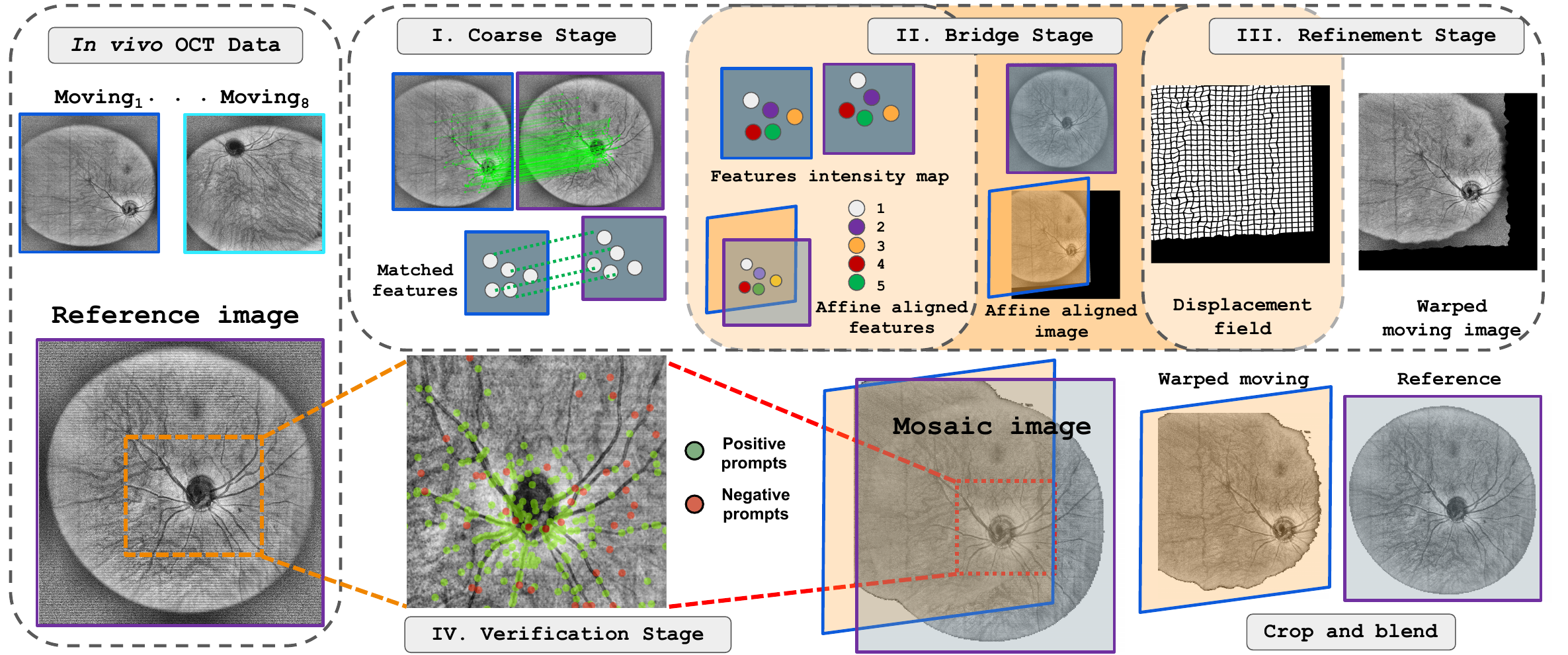}   
\caption{Four-stage OCT  mosaicking pipeline.  Stage I displays feature matching: the moving image in blue and the reference in purple, and 5 sample matched features. Stage II shows the features-intensity affine transform applied to the moving image. In Stage III, we use deformable registration to refine the alignment. In Stage IV, the final mosaic image is evaluated by SAM.} 
\label{fig:pipeline}
\end{figure*}

In this paper, we focus on OCT \textit{en face} image mosaicking. We present an innovative pipeline that bridges the two categories: learning-based feature matching \cite{lindenberger2023lightglue} and pixel-based non-rigid registration \cite{avants2008symmetric}. Our methodology is inspired by the state-of-the-art (SOTA) work in feature matching, leveraging pretrained neural networks to enhance the robustness in challenging scenarios, and is built upon the well-established image registration toolbox, Advanced Normalization Tools (ANTs) \cite{avants2009advanced}. To this end we propose a novel feature-intensity matching strategy, which serves as a bridge between these two approaches. We validate our method on a noisy multi-field dataset with prominent deformation between images, and we also evaluate accuracy and efficiency in a large public dataset.

A common problem with medical image mosaicking is the lack of a standardized approach for evaluating results in the lack of dense ground truth \cite{bano2020deep}. We propose a novel evaluation tool leveraging the Segment Anything (SAM) \cite{kirillov2023segment} foundational model with prompt engineering \cite{li2023assessing}. Integrating matching features as prompts within the mosaicking pipeline allows us to conduct automated validation even when ground truth is unavailable. 

Our novel contributions are:

\begin{itemize}
    \item We propose a novel image mosaicking pipeline that combines the features-based method for global alignment and the pixel-based method for local refinement, enhancing the robustness and accuracy of the process.
    \item We rigorously validate our approach, demonstrating its capability to produce high-quality OCT panoramic images even in the presence of substantial noise and deformation.
    \item We explore foundational models as a tool to assess the performance of medical image mosaicking without supervision.
\end{itemize}

\begin{figure}[t]
      \centering
      \centerline{\includegraphics[width=\linewidth]{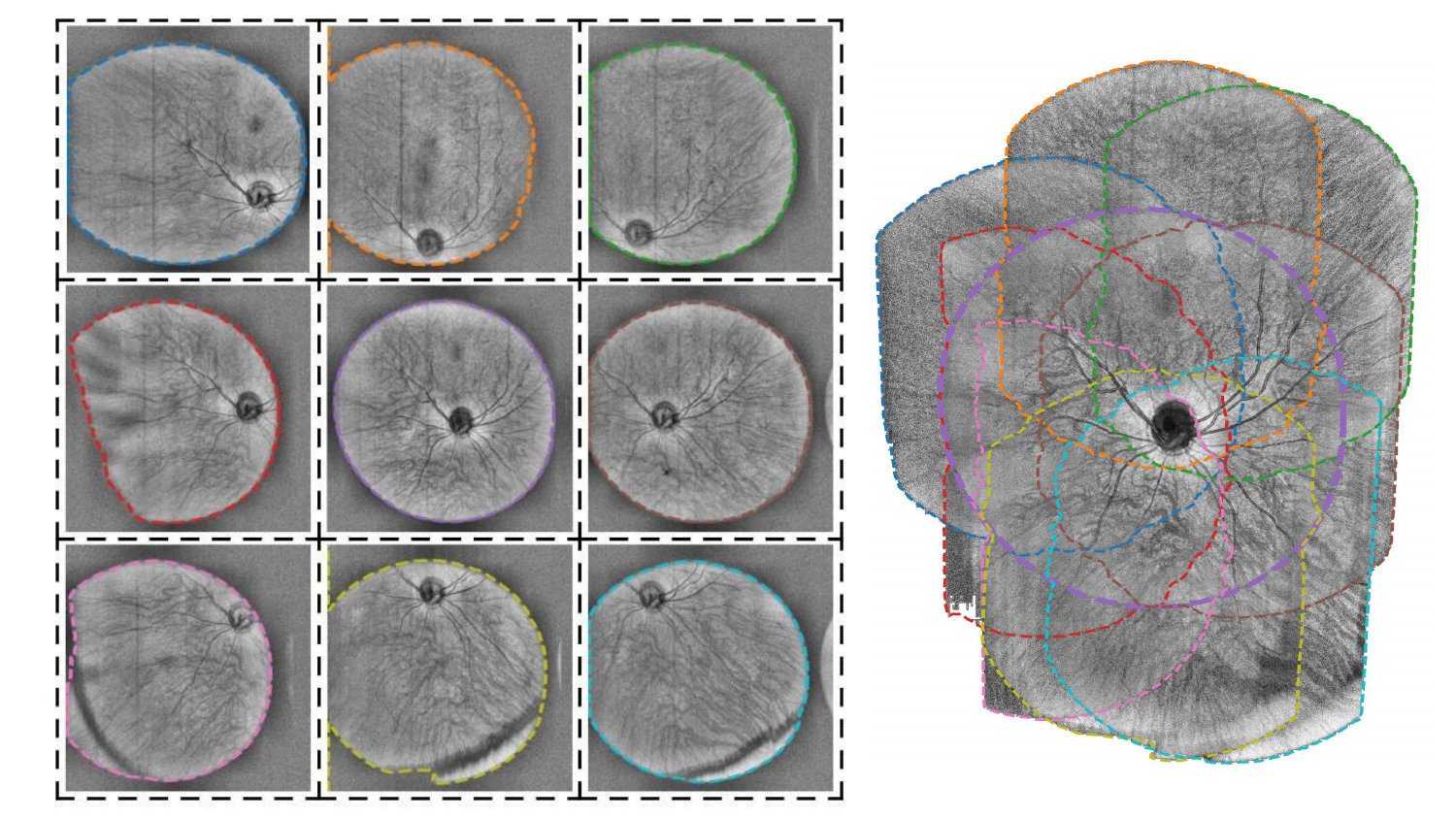}}
      \caption{\textbf{Left:} Original images with the central (purple) field serving as the fixed image. \textbf{Right:} Mosaicking result with a larger FoV. The contours of each sub-field are color-coded.}% A FoV exceeding 90 degrees is successfully achieved.}
    \label{fig:mosaic}
    \end{figure}
    
\section{Methods}
\label{sec:method}

    Our proposed pipeline, illustrated in Figure \ref{fig:pipeline}, adopts a coarse-to-fine strategy and is divided into four stages. It is designed to process 2D images of OCT \textit{en-face} projection images, producing a mosaic image with a wider field of view as depicted in Figure \ref{fig:mosaic}. The process begins with coarse feature-based registration, where affine transformation matches feature intensities across images. Subsequently, the refinement stage builds upon this coarse alignment, focusing on fine-tuning local variations through deformable image registration. Finally, we utilize feature points identified in the coarse stage to verify and assess the mosaicking quality.
    
    \noindent\textbf{Coarse Stage: Feature-based alignment.} This stage involves feature detection and matching. For detecting features, we employ SuperPoint \cite{detone2018superpoint}, which is optimized through self-supervised training on a synthetic dataset for enhanced keypoint detection. Compared to traditional methods such as SIFT \cite{lowe2004distinctive}, SURF \cite{bay2008speeded}, and ORB \cite{rublee2011orb}, SuperPoint captures a larger number of high-quality features. We use LightGlue \cite{lindenberger2023lightglue} for matching these feature points. As an advancement over SuperGlue \cite{sarlin2020superglue}, LightGlue improves matching accuracy with relative positional encoding and an adaptive pruning design. Moreover, LightGlue demonstrates superior robustness and efficiency when compared to conventional matching methods like brute force, KNN, and FLANN \cite{lindenberger2023lightglue}. We also compare the performance with the learning-based dense-feature matching method LoFTR \cite{sun2021loftr}, which matches features directly at the pixel level, skipping the feature detection step.
        
    \noindent\textbf{Bridge Stage: Affine registration for features.} In this stage, we introduce a new strategy for registering matching points between images. Historically, Iterative Closest Point (ICP) \cite{arun1987least} and its variants have been common for point registration. However, these methods often depend heavily on initial conditions to avoid getting stuck in local optima. Another frequent method is RANSAC-based \cite{fischler1981random} homography estimation, but it can cause shearing distortions due to uneven distribution of local features \cite{liu2022content}, leading to misalignment. 
    
    We propose to create feature images by assigning unique intensity values to each matching feature pair in both images. This allows us to employ a direct intensity-based affine image registration that uses a global correlation metric to find the best global transformation. We intentionally restrict the degrees of freedom compared with homography estimation to minimize distortions from noisy or isolated features.
    
    \noindent\textbf{Refinement Stage: Pixel-based deformable registration.} In this stage, we refine pixel-wise similarity through direct registration. Starting with the affine transform results from the previous stage, we use the ANTs SyN (Symmetric Normalization) algorithm \cite{avants2008symmetric} to estimate the diffeomorphic deformation field that minimizes the cross-correlation metric. We use a 5-scale registration scheme by setting appropriate shrink factors and sigma smoothing.
    
    \noindent\textbf{Verification Stage: Features-prompts for SAM segmentation.} Inspired by previous work in fetoscopic mosaicking \cite{bano2020deep}, which emphasizes the importance of maintaining vessel features post-mosaicking, we devise a strategy to evaluate how well the mosaic images preserve vessel features. However, segmenting OCT vascular structures is highly challenging \cite{hu2023deep}, mainly due to the difficulty in obtaining ground truth for small, densely clustered capillary vessels. To address this challenge, we utilize the foundation model SAM \cite{kirillov2023segment}. If the SAM segmentation of the reference image is the same as the SAM segmentation of the mosaic image (in the overlapping region), we consider the mosaic successful. In contrast, errors in alignment will create ghost vessels in the mosaic image, which will cause over-segmentation in the SAM results. We create the SAM segmentation by using feature points from SuperPoint as prompts: we use the features that are matched by LightGlue as positive prompts, and the features that are assigned lower confidence by SuperPoint as negative prompts. %This allows us to assess how well vessel features are preserved in the mosaic image, without needing additional supervision. 

\section{Experiments and Results}
\label{sec:experiments}
\subsection{Datasets}

\noindent\textbf{In-house 9-field dataset:}
We acquired images from two healthy control subjects using our  Spectrally Encoded Coherence Tomography and Reflectometry (SECTR) system \cite{el2018spectrally} under a protocol approved by the Vanderbilt University Medical Center IRB. This included nine high-resolution OCT sub-field volumes (2560 × 2000 × 1400 pixels) per subject, covering over 45 degrees of field of view (FoV). 2D en-face projections were computed by depth-averaging OCT volumes without using denoising processes, to test our pipeline's noise robustness. Motion correction and CLAHE were applied for image intensity standardization.

\begin{figure}[t]
  \centering
  \centerline{\includegraphics[width=\linewidth]{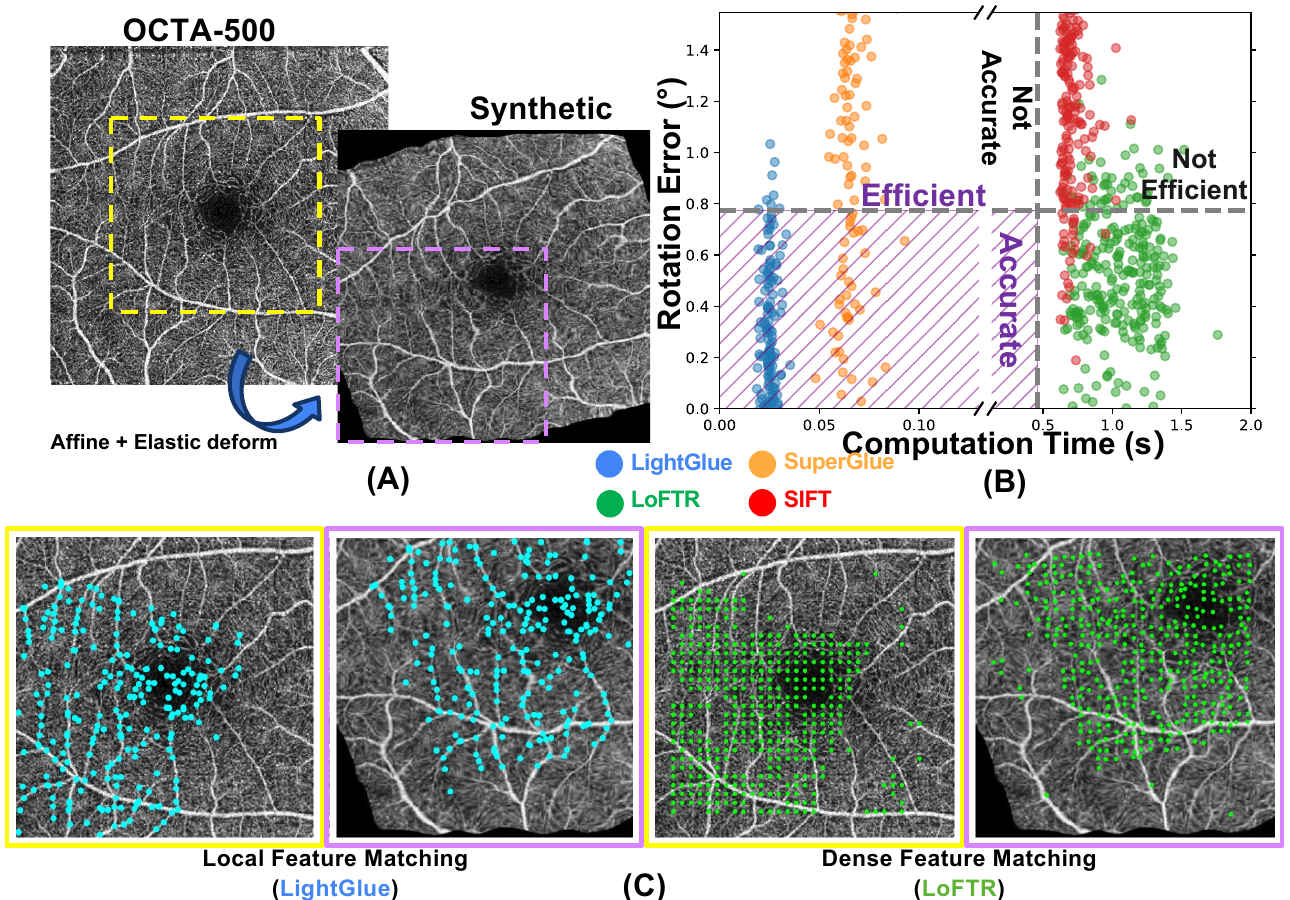}}
  \caption{\textbf{(A)} OCTA-500 Synthetic dataset: known random affine transform and elastic deformation were applied to the cropped fields (2 fields shown in dashed lines). \textbf{(B)} Rotation estimation error (degrees) and computation time (seconds) of each feature matching method. Each dot represents an individual subject's result, color-coded by methods. The highlighted quadrant shows LightGlue is the most desirable choice, combining high accuracy and efficiency. \textbf{(C)} Qualitative feature point comparison. LightGlue focuses on sparse but representative points, whereas LoFTR yields a dense grid of features that may not be as informative.}
  \label{fig:synthetic}
\end{figure}

\begin{figure*}[ht]
      \centering
      \centerline{\includegraphics[width=1\linewidth]{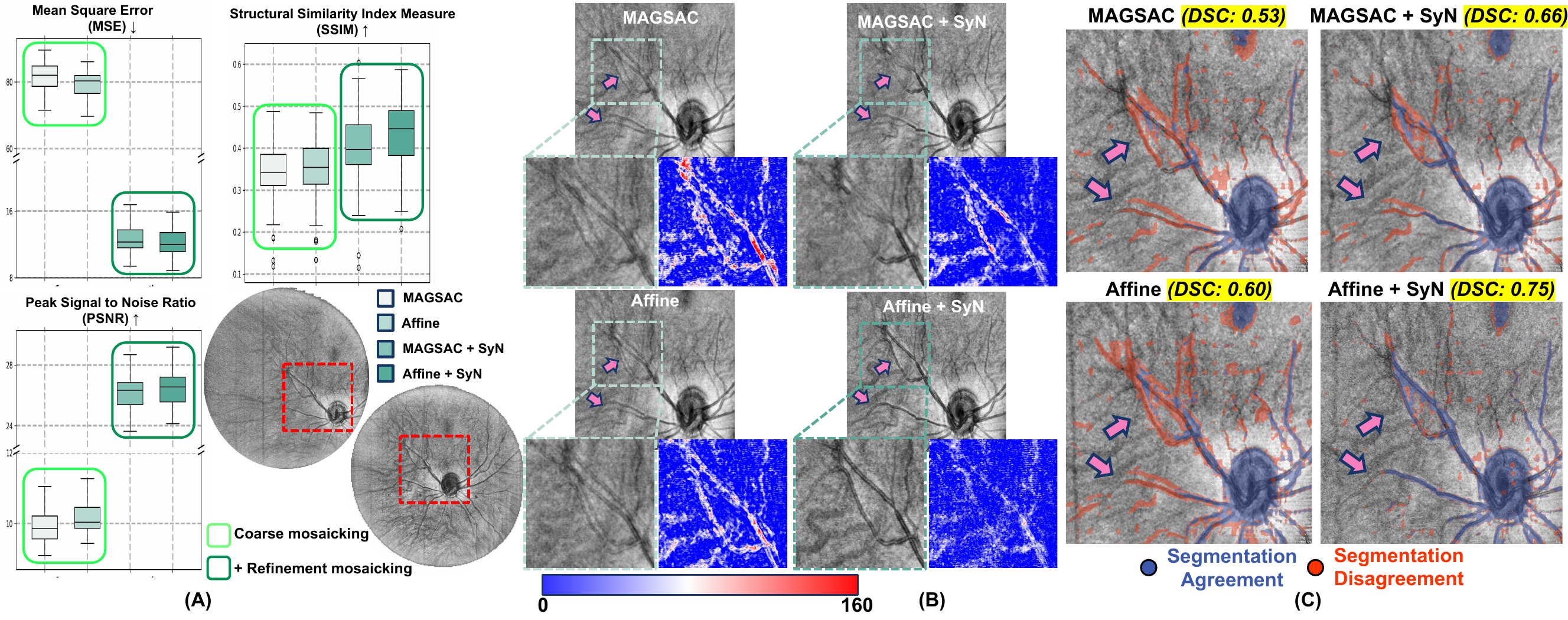}}
    \caption{\textbf{(A)} Quantitative comparison in the 9-field dataset: Light colors (MAGSAC and Affine) represent coarse stage results, while dark colors (MAGSAC+SyN and Affine+SyN) represent the refinement stage. Our method (Affine+SyN) outperformed MAGSAC in all three metrics at both stages. \textbf{(B)} Qualitative comparison, with pink arrows highlighting noteworthy differences. The intensity difference between the mosaic and reference images ($\Delta = |I_{ref}-I_{mosaic}|$) in the dashed region is represented using a blue-to-red color map.  \textbf{(C)} SAM segmentation from the mosaic is compared against the reference image segmentation. Blue areas indicate segmentation agreement, while orange regions reveal  discrepancies. Dice score for each method is reported.}
    \label{fig:invivo}
\end{figure*}

\noindent\textbf{Synthetic public OCTA-500 dataset:}
    To assess our pipeline in a larger dataset, we used the publicly available OCTA/OCT dataset, OCTA-500 \cite{li2020octa}, which includes 300 human OCTA en-face projections (400 $\times$ 400 pixels). We artificially created sub-fields for mosaicking by cropping ROIs from the large FoV image and artificially distorting them, as seen in Figure \ref{fig:synthetic} (A). We thus created a 5-sub-field mosaic with random affine transformations (rotation range [-20, 20] degrees, translation range [-10, 10] pixels) and elastic displacement fields (magnitude range [-5, 5] pixels) to simulate real-world mosaicking conditions. %These images were processed to form corner patches with varied translation magnitudes (5, 7, and 9 pixels) for further mosaicking tests.
    These known transformations are used for quantitative evaluation of our mosaicking pipeline.

\subsection{Experiment I: Feature selection and matching}

    We first compared our adopted learning-based local feature matching method (LightGlue) against other feature matching methods. The primary metric for this comparison was the estimated rotation angle error $\Delta$, calculated on the synthetic OCTA-500 dataset. Our findings revealed that LightGlue ($\Delta = 0.52 \pm 0.24$) and LoFTR ($\Delta = 0.47 \pm 0.35$) exhibited nearly identical performance ($p > 0.3$ in paired t-test) and significantly  outperformed both SuperGlue ($\Delta = 1.21 \pm 0.61$, $p < 0.005$), and SIFT ($\Delta = 1.75 \pm 0.87$, $p < 0.005$). Additionally, we assessed these methods in terms of time consumption, a vital consideration in medical imaging applications, as illustrated in Figure \ref{fig:synthetic} (B). LightGlue emerged as the preferred choice, offering an optimal balance between accuracy and efficiency. Figure \ref{fig:synthetic} (C) demonstrates that LightGlue identified more relevant matching features (vessel structure) compared to LoFTR. LoFTR produces a dense collection of matching features, but these appear less informative for subsequent steps. Furthermore, matching based on sparser features is computationally more efficient.    

\subsection{Experiment II: Affine alignment for features}
\label{sec_exp2}
    We compared our proposed affine alignment strategy against the SOTA RANSAC-based MAGSAC method \cite{barath2019magsac} in both coarse and refinement  stages using the in-house 9-field  dataset. While there are only two subjects in this dataset, we considered each pair of images in the set of 9 sub-fields, resulting in 2 subjects $\times$ 9 fixed sub-fields $\times$ 8 moving sub-fields = 144 pairs. Given the lack of ground truth in this dataset, we assessed the alignment discrepancies using three metrics: Root Mean Square Error (RMSE), Mean Structural Similarity Index (SSIM), and Peak Signal to Noise Ratio (PSNR), as depicted in Figure \ref{fig:invivo} (A). Our affine alignment method outperformed MAGSAC in all three metrics at both stages, with improvements reaching statistical significance ($p_{MSE}<1e-5, p_{SSIM}<1e-4, p_{PSNR}<1e-5$) in paired t-tests between Affine+SyN vs.\ MAGSAC.

    Figure \ref{fig:invivo} (B) showcases a visual comparison of mosaicking results using different methods for a given pair of sub-fields. The zoomed-in panels show the intensity difference map between the mosaic and reference images as a color map, within the overlapping regions. We observe pronounced registration errors in the MAGSAC result, evidenced by the `ghost` vessels such as those highlighted by the pink arrows. Our affine method mitigates these artifacts to some extent. After the refinement stage, the MAGSAC method still has notable misalignment artifacts, whereas our method exhibits well-aligned vessels after the refinement stage.
    
\subsection{Experiment III: Unsupervised  evaluation with SAM}
    We use SAM to evaluate the mosaic quality in an unsupervised manner, given the lack of ground truth data in medical image mosaicking. This stage leverages the SuperPoint feature points identified in both the reference and moving images in Stage I. We designated the feature points matched by LightGlue as positive prompts, and the feature points in the reference image with low confidence scores ($\leq 0.0009$) as negative prompts. These feature prompts were inputted into the pre-trained SAM model to segment both the mosaic and the reference images, with no additional fine-tuning. We compared the segmentation results  from the mosaic image and from the reference image. This comparison is visually represented in Figure \ref{fig:invivo} (C), where the orange regions indicate mismatches in segmentation, highlighting considerable discrepancies in ghost vessel regions identified in the previous experiment. We also report the Dice score between the SAM vessel segmentation results pre- and post-mosaicking. These Dice results support our qualitative hypothesis: the fine-tuning stage improves mosaicking outcomes, with our adapted method achieving the highest Dice score (Figure \ref{fig:invivo} (C), $DSC = 0.75$). We believe that this verification process holds potential for enhancing future comprehensive evaluations of image mosaicking methods, leveraging segmentation comparison metrics such as the Dice score, average surface distance (ASD), and the 95-percent Hausdorff distance.

\section{Discussion and Conclusions}
\label{sec:discussion}

In summary, our paper introduces a novel image mosaicking pipeline specifically designed for handling noisy and deformed OCT/OCTA images. We explored how to leverage advanced learning-based local feature matching methods for feature registration with an intensity-based approach. We have demonstrated superior performance on both in-house and public synthetic datasets. Additionally, our innovative use of feature prompts for unsupervised verification presents a new way to assess the quality of image mosaicking.

\noindent\textbf{Acknowledgements.}
This work is supported, in part, by the NIH grants R01-EY033969, R01-EY030490 and R01-EY031769.

\noindent\textbf{Compliance with Ethical Standards.}
%\section{Compliance with Ethical Standards}
This research study was conducted retrospectively using human subject data made available in open access by OCT-500. Ethical approval was not required as confirmed by the license attached with the open access data. The in-house data was acquired in a study approved by the Vanderbilt University Institutional Review Board (IRB No.\ 202230).

% Below is an example of how to insert images. Delete the ``\vspace'' line,
% uncomment the preceding line ``\centerline...'' and replace ``imageX.ps''
% with a suitable PostScript file name.
% -------------------------------------------------------------------------
% \begin{figure}[htb]

% \begin{minipage}[b]{1.0\linewidth}
%   \centering
%   \centerline{\includegraphics[width=8.5cm]{example-image}}
% %  \vspace{2.0cm}
%   \centerline{(a) Result 1}\medskip
% \end{minipage}
% %
% \begin{minipage}[b]{.48\linewidth}
%   \centering
%   \centerline{\includegraphics[width=4.0cm]{example-image}}
% %  \vspace{1.5cm}
%   \centerline{(b) Results 3}\medskip
% \end{minipage}
% \hfill
% \begin{minipage}[b]{0.48\linewidth}
%   \centering
%   \centerline{\includegraphics[width=4.0cm]{example-image}}
% %  \vspace{1.5cm}
%   \centerline{(c) Result 4}\medskip
% \end{minipage}
% %
% \caption{Example of placing a figure with experimental results.}
% \label{fig:res}
% %
% \end{figure}

% \vfill
% \pagebreak

\bibliographystyle{IEEEbib}
\bibliography{refs}

\end{document}